\documentclass[pdflatex,sn-basic]{sn-jnl}

\usepackage{url}
\usepackage{verbatim}
\usepackage{listings}
\usepackage{multirow}
\usepackage{hyperref}
\usepackage{algpseudocode}
\usepackage{enumitem}
\usepackage{algorithm}
\usepackage{amsmath,amssymb,amsfonts}%
\usepackage{amsthm}%
\usepackage{mathrsfs}%
\usepackage[title]{appendix}%
\usepackage{xcolor}%
\usepackage{textcomp}%
\usepackage{manyfoot}%
\usepackage{booktabs}%
\newcommand\tool[0]{GIDroid}

\begin{document}

% The paper headers
\title[Article Title]{Multi-Objective Improvement of Android Applications}

\author*[1]{\fnm{James}, \sur{Callan}}\email{james.callan.19@ucl.ac.uk}
\author*[1]{\fnm{Justyna} \sur{Petke}}\email{j.petke@ucl.ac.uk}
\affil*[1]{\orgdiv{Computer Science Department}, \orgname{University College London}, \orgaddress{\street{Gower Street}, \city{London}, \postcode{WC1E 6BT}, \state{Greater London}, \country{United Kingdom}}}

\abstract{
Non-functional properties, such as runtime or memory use, are important to mobile app users and developers, as they affect user experience.
Previous work on automated improvement of non-functional properties in mobile apps failed to address the inherent trade-offs between such properties.

We propose a practical approach and the first open-source tool, \cite{repo}, for multi-objective automated improvement of Android apps.
In particular, we use Genetic improvement, a search-based technique that navigates the space of software variants to find improved software.
We use a simulation-based testing framework to greatly improve the speed of search.
\tool~contains three state-of-the-art multi-objective algorithms, and two new mutation operators, which cache the results of method calls.

Genetic improvement relies on testing to validate patches.
Previous work showed that tests in open-source Android applications are scarce.
We thus wrote tests for 21 versions of 7 Android apps, creating a new benchmark for performance improvements.

We used \tool~to improve versions of mobile apps where developers had previously found improvements to runtime, memory, and bandwidth use.
Our technique automatically re-discovers 64\% of existing improvements.
We then applied our approach to current versions of software in which there were no known improvements.
We were able to improve execution time by up to 35\%, and memory use by up to 33\% in these apps.}

\maketitle

\section{Introduction}
\label{sec:Intro}

Android applications (or apps for short) are one of the most widely used types of software~(\cite{kemp_2022}).
They are designed for direct user interaction, with the main entry point for the software being its UI components.
Due to the small size of Android devices (phones and tablets) compared to traditional desktop devices, their hardware capabilities are naturally limited.
These two factors result in non-functional properties being especially important to both users and developers.
In fact, non-functional properties are so important to Android users that 1/3 of instances of users abandoning applications and 59\% of bad reviews were due to poor performance~(\cite{userCountries,inukollu2014factors}).\looseness=-1

\cite{9397392}'s survey on Android performance optimizations lists several approaches for improving non-functional properties of Android apps.
These include prefetching online resources to avoid having to wait for them when they are needed~(\cite{prefetchAds,DBLP:journals/imwut/BaumannS17}) and offloading computation onto remote servers which are faster than the mobile device~(\cite{offloadComs}).
Offloading, however, requires external server infrastructure to be set up and applications to be re-engineered to be utilised.
Prefetching is only applicable to areas of applications that interact with the network.
Other approaches~(\cite{9397392}) include anti-pattern detection, which requires manual implementation, and refactorings, which are limited to specific code fragments.
We argue that an approach that does not require external resources and 
is more easily applicable to all applications regardless of type and structure would make developers more likely to adopt it. 

Whilst existing approaches for automated improvement of Android apps are capable of improving multiple properties simultaneously, e.g., by removing unnecessary computation reducing runtime and energy use, in most cases such correlations have not been considered~(\cite{9397392}).
%For example, when reducing the energy usage of applications using offloading, responsiveness improvements are simply a positive side-effect.
Moreover, single-objective improvements can have negative effects on other properties.
For example, during prefetching, the resource which is prefetched must be stored, which might result in higher memory use.
To get the full picture of how an application is affected by an improvement, properties other than those that are direct targets for improvement should be considered.
\cite{9397392} reveals only one work that applies multi-objective optimization to non-functional properties of Android apps. 
\cite{DBLP:journals/tse/MoralesSKCA18} consider energy consumption and the number of anti-patterns.
Although the authors release their framework, it is not open-source and requires external hardware for energy measurements.

Rather than targeting specific features or resources, we aim to find source code transformations.
There have been a few attempts to find Android app performance optimizations with source code transformations so far.
Lin et al. proposed two approaches, Asynchronizer~(\cite{DBLP:conf/sigsoft/LinRD14}) and AsyncDroid~(\cite{DBLP:conf/kbse/LinOD15}), for refactoring code to be executed asynchronously. 
However, both of these approaches require developers to identify the particular lines of code which they want to execute asynchronously and there has been no work to show the actual impact of these refactorings on performance.
\cite{DBLP:conf/issta/LyuLH18} propose an approach that moves costly database operations out of loops.
Whilst this approach can improve performance, it is only applicable to methods that access databases inside loops.

The only tools for Android app performance improvement, which are both available and generally applicable to Android source code, are linters~(\cite{DBLP:conf/kbse/HabchiBR18}). 
Linters contain rules which aim to identify areas of code that may cause performance issues, leaving to app developers the decisions to implement suggested changes. 
However, their use comes with challenges~(\cite{DBLP:conf/kbse/HabchiBR18}), including dealing with false positives.

In order to find patches to source code, we propose to use Genetic Improvement. GI is a search-based technique that uses meta-heuristics to perform a guided search over software patches, to find those that improve a given software property.
GI makes changes to source code and thus can be applied to a wide range of software types.
GI has been used to improve many different properties of software, including runtime~(\cite{DBLP:conf/gecco/LangdonLPH15,DBLP:conf/ssbse/PetkeLH13}), memory~(\cite{DBLP:conf/ssbse/BasiosLWKB17,DBLP:conf/gecco/WuWHJK15}), and energy consumption~(\cite{DBLP:conf/gecco/BrucePH15,DBLP:conf/ssbse/BurlesBBKSV15}).

%Not only has GI been used to improve properties individually, but also 
%<<<<<<< HEAD
%=======
%GI has also been used to improve multiple properties at once in the desktop domain~\cite{DBLP:conf/kbse/MesecanB0CP22,DBLP:conf/gecco/WuWHJK15,DBLP:conf/ssbse/BasiosLWKB17, DBLP:conf/gecco/WuWHJK15,DBLP:conf/ssbse/CallanP22}.
%>>>>>>> refs/remotes/origin/master
%Extending GI to improve multiple properties can be accomplished by swapping out these single-objective algorithms with multi-objective ones.

Extending GI to improve multiple properties can be accomplished by swapping out these single-objective algorithms with multi-objective ones.
This allows us to consider patches that find trade-offs between various properties, rather than just those which improve one, without consideration of the impact on others. 
We can thus provide a choice to developers between different versions of source code, showing different trade-offs.
Nevertheless, only a few works explore the potential of multi-objective GI and only in the desktop domain~(\cite{DBLP:conf/kbse/MesecanB0CP22,DBLP:conf/gecco/WuWHJK15}).

GI has been applied to Android applications a handful of times. 
Callan and Petke attempted to improve the frame rate of Android apps with GI, however, were unsuccessful~(\cite{DBLP:conf/ssbse/CallanP21}).
In another work, \cite{DBLP:conf/icse/CallanP22} were able to find improvements to the navigation responsiveness of Android apps.
\cite{DBLP:conf/gecco/BokhariBA017} improved the energy consumption of Android apps, with a type of GI known as deep parameter optimization.
To the best of our knowledge, no GI work so far has attempted to improve and find trade-offs between multiple properties of Android apps, and no approach has attempted to improve either the memory consumption or bandwidth usage of Android apps, which we target in this work.

Previous work on applying GI in the Android domain revealed several practical challenges: 1) due to the complexity of the Android build system and significant use of UI elements, a minor change usually requires a time-costly process of installation on the actual device for testing 2) tests themselves are scarce, and 3) performance fitness measurements used in the desktop domain are not accurate enough to witness performance issues in Android apps, yet users deem wait time of just 150ms as `laggy'~(\cite{tolia2006quantifying}).
We overcome these challenges. 
We utilise the Robolectric testing library~(\cite{robolectric}) which mimics UI behaviour, allowing for quick unit testing, without need for installation on an actual mobile or tablet device. 
This simulation-based approach provides us with means of utilising performance measurement tools unavailable on Android devices.\looseness=-1

In order to validate our proposed approach, we created a tool, \cite{repo}, for running multi-objective (MO) GI on Android applications.
We provide three fitness functions, to improve runtime, memory, and bandwidth use.
\tool~contains three MO algorithms (NSGA-II~(\cite{DBLP:conf/ppsn/DebAPM00}), NSGA-III~(\cite{DBLP:journals/tec/DebJ14}), and SPEA2~(\cite{DBLP:conf/ppsn/KimHMW04})). 
Based on work by~\cite{DBLP:journals/ese/CallanKPS22}, who mined non-functional improvements made by Android developers, we implement in \tool~two novel mutation operators, specifically designed to mimic human-made edits.
These cache results of repeated calls, aiming to save memory use.

When using GI, we validate the patches that we generate using the program's test suite and validate the best-improving final patches manually. 
This ensures that our patches do not disrupt the functionality of the program.
However, most open-source Android applications do not have test suites, and those that do are limited, achieving a median line coverage of 23\%~(\cite{DBLP:conf/iwpc/PecorelliCFLP20}).
This meant that we had to create tests for all the benchmarks on which we ran GI.\footnote{
At the time of our experiments, none of the automated test generation tools for Android were compatible with latest Android software, thus we had to manually create tests to evaluate our MO-GI approach. 
}

We selected Android apps that contain real-world non-functional-property-improving commits, in order to see if \tool~can re-discover changes made by Android developers.
Moreover, we use the latest versions of these applications, to see if we can find as-yet-undiscovered improvements.
Overall, we created a benchmark of 21 versions of 7 Android apps, which we open source for future work.

\tool~was able to find patches that improve execution time by up to 35\%, and memory usage by up to 65\%. 
Unfortunately, no improvements to bandwidth use were found. 
Such improvements are within \tool's search space, which leaves room for future work for more effective search strategies.

To sum up, we present the following novel contributions:
\begin{enumerate}[topsep=5pt]
    \item An open-source, simulation-based tool, \tool~(\cite{repo}), for automated multi-objective improvement of Android applications' runtime, memory, and bandwidth use.
    \item A benchmark of 21 versions of 7 Android applications, including tests, for future work on performance improvement in the Android domain.
    \item An evaluation of the effectiveness of 3 multi-objective genetic improvement algorithms at improving runtime, memory use, and bandwidth of Android applications. No GI work has targeted 3 properties before.
    \item A comparison between both multi- and single-objective genetic improvement approaches for automated optimization of Android applications.
    \item An empirical comparison of our multi-objective GI-based approach for Android application performance improvement with state-of-the-art linters.
\end{enumerate}

The rest of this paper is structured as follows:
Section~\ref{sec:relWork} describes related work;
Section~\ref{sec:Background} presents an introduction to genetic improvement and multi-objective optimization;
Section~\ref{sec:GI} presents challenges of applying GI to the Android domain and our proposed framework that overcomes these challenges;
Section~\ref{sec:RQs} presents research questions we aim to answer to evaluate our approach, with 
Section~\ref{sec:Meth} outlining our methodology;
Section~\ref{sec:Res} presents our results,
with threats to validity presented in Section~\ref{sec:Threats};
Section~\ref{sec:Cons} concluding.

\section{Android App Performance Optimization}
\label{sec:relWork}

A number of approaches have been proposed for improving the performance of Android applications.
\cite{9397392}'s survey on this topic presents the following code-level approaches:

\textbf{Prefetching:} Network resources are fetched before they are needed by the application and stored locally~(\cite{prefetchAds, DBLP:journals/imwut/BaumannS17}). 
    When the application needs said resources, it can get them without having to wait for a lengthy network transaction, making the application more responsive.
    Prefetching can lead to increased memory and storage usage, and lead to the app not having the most up-to-date version of a particular resource.
    Prefetching can only optimize parts of applications that utilise network resources.

\textbf{Anti-patterns:} Approaches that detect patterns in source code that indicate performance defects, for example, repeated expensive memory access operations inside for-loops~(\cite{6606602}). 
The only tools which are both available and generally applicable to the source code of Android apps are linters~(\cite{androidLint}), PMD~(\cite{pmd}), and FindBugs~(\cite{findbugs}).
These tools have performance rules which aim to identify areas of code that may cause performance issues.
However, often these warnings are false-positives~(\cite{DBLP:conf/kbse/HabchiBR18}).
The developer must then manually fix the issues.
Existing Android linters do not provide any information on the impact of fixing the issues they detect.

\textbf{Refactoring:} Refactoring approaches aim to modify the source code of the application to be more performant.
In \cite{DBLP:conf/sigsoft/LinRD14,DBLP:conf/kbse/LinOD15}'s work  applications were refactored to execute code asynchronously, making them execute more quickly.
These approaches require developers to identify each line of code that they wish to execute asynchronously and there is no indication of the actual impact on performance of these changes.
\cite{DBLP:conf/issta/LyuLH18} propose to move database operations 
 out of loops. 
However, this is only applicable to limited areas of code that contain such database calls.

\textbf{Offloading:} This approach aims to perform the most costly computation on external servers, rather than Android devices~(\cite{APPS,cloneCloud,APOff,preemptOff,offloadComs}). 
    This has the benefit of reducing the amount of energy used by the application, extending the device's battery life, and speeding up the computation to make the app more responsive.
    Offloading requires external hardware to function, which may not always be suitable.

\textbf{Programming Languages:} In the Android environment, a number of different programming languages are available to developers.
    The majority of Android apps are written in either Java or Kotlin, which usually compile to JVM bytecode.
This bytecode is then (optionally) obfuscated and recompiled into dex code. 
    This allows Java and Kotlin APIs to be used across both languages interchangeably and some applications even use a mixture of both languages.
    There is little performance difference between the two languages~(\cite{DBLP:journals/corr/abs-2103-09728}).
    C/C++ can also be used to write native code.
Such code is generally faster than the Java/Kotlin code and thus can be used to find performance improvements. 
However, changing a programming language can be time-consuming, with no upfront knowledge of the magnitude of possible performance gains.\looseness=-1

The above works have all proved useful, but they either do not perform fully automatic improvement~(\cite{DBLP:conf/sigsoft/LinRD14,DBLP:conf/kbse/LinOD15,DBLP:conf/kbse/HabchiBR18,findbugs,androidLint}), are only applicable to specific areas of code~(\cite{DBLP:conf/issta/LyuLH18,prefetchAds, DBLP:journals/imwut/BaumannS17}), or require external infrastructure~(\cite{APPS,cloneCloud,APOff,preemptOff,offloadComs}).\looseness=-1

Given the shortcomings of the above-mentioned approaches, we propose to use multi-objective GI to improve several software properties in the Android domain.
By using GI, we will be able to apply our approach to any source code and will not be limited to only improving code using certain patterns.
GI is fully automated.
Developers will only have to review the patches produced by GI once the process is finished to ensure that they do not have unintended side effects.
Such patches would thus undergo a standard code review process.
Furthermore, GI does not require the setup of any external infrastructure to achieve optimisations and can be performed in the local development environment of the application developer.
We illustrate this in Table~\ref{tab:relWork}. 
We aim to find multiple patches, which may find trade-offs between different properties that can allow developers to choose the best patches for their particular needs, and be fully aware of the consequences that a particular patch will have on other properties.
We note that prefetching, offloading, and others are complementary to GI, and could still offer benefits to applications that have been optimized using GI.

\begin{table}
     \centering
    \caption{Comparison of existing strategies for improvement of non-functional properties of Android apps with our tool -- \tool.}
    \addtolength{\tabcolsep}{-2pt}
    \begin{tabular}{llccc}
    \hline
    Work & Properties & Source & Fully & Trade-offs\\
     &  & Code & Automatic & Considered\\
    \hline
    Prefetching%~(\cite{prefetchAds, DBLP:journals/imwut/BaumannS17} )
    & Runtime & x & \checkmark & x\\
    Anti-Patterns%~(\cite{androidLint, pmd, findbugs, DBLP:conf/kbse/HabchiBR18})
    &  Runtime, Memory & \checkmark & x & x\\
    Refactoring - Asynchronous%~(\cite{DBLP:conf/sigsoft/LinRD14,DBLP:conf/kbse/LinOD15})
    & Runtime & \checkmark & x & x\\
    Refactoring - Database Loops%~(\cite{DBLP:conf/issta/LyuLH18}) 
    & Runtime & \checkmark & \checkmark & x\\
    Offloading%~(\cite{APPS,cloneCloud,APOff,preemptOff,offloadComs})
    & Runtime, Energy Use & x & \checkmark & x \\
    \hline
    \textbf{\tool} & Runtime, Memory, Bandwidth & \checkmark & \checkmark & \checkmark \\
    \hline
    \end{tabular}
    \label{tab:relWork}
\end{table}

\section{Background}
\label{sec:Background}

Before we outline our proposed framework for automatic performance improvement of Android applications, we first provide a short introduction to Genetic Improvement (GI) and Multi-Objective (MO) optimization.
%Next, we discuss practical challenges related to applying GI on Android.
%Finally, we describe how we overcome the challenges and present our approach for MO-GI for the mobile domain.

\subsection{Genetic Improvement}
Genetic Improvement (GI)~(\cite{DBLP:journals/tec/PetkeHHLWW18}) is a search-based software engineering technique that utilises search to iterate over different versions of software in order to find improved program variants.
These improvements can be bug repairs or improvements to non-functional properties like execution time or memory use.
GI has already proven useful for improvement of traditional software, fixing bugs during the development of commercial software~(\cite{DBLP:conf/gecco/HaraldssonWBS17}), improving the execution time of large bioinformatics software~(\cite{DBLP:conf/gecco/LangdonLPH15}), improving compiler optimizations~(\cite{DBLP:conf/gecco/LiPSRB22}), and more~(\cite{DBLP:journals/tec/PetkeHHLWW18}).
 
Each program in GI is represented as a patch to existing software.
Patches are constructed from a set of edits to code, i.e., mutations, which describe modifications to the program being improved.
The most common mutation operators used in previous work have been: {\sc delete}, {\sc copy}, and {\sc replace}.
These operations can be applied at the level of lines of source code, bytecode, or other.
The vast majority of GI work operates at statement-level, applying mutation operators to nodes of an abstract syntax tree~(AST).\looseness=-1

Each GI patch is applied to the original software for evaluation, measured using a fitness function.
In the case of program repair, this fitness can be the number of passing tests, and for execution time improvement it could be the time taken by the tests.
This fitness measurement is used to guide search through the landscape of patches to find improved software variants.
Traditionally, genetic programming has been used for this purpose, though other search techniques, such as local search, have also proven effective~(\cite{DBLP:journals/tec/BlotP21}).

Although there is a lot of literature on the improvement of traditional software using GI, little is known about how the technique would fare in the mobile domain.
Initial approaches have shown mixed results, with none trying to optimize multiple properties.
\cite{DBLP:conf/gecco/BokhariBA017} were able to reduce the energy consumption of Android apps, using deep parameter optimization, i.e., mutating parameters within source code, not exposed to the user.
\cite{DBLP:conf/icse/CallanP22} were able to reduce the time taken to move between Activities, the main UI components, of Android apps.
However, when attempting to improve the frame rate of Android apps, Callan and Petke did not find improving patches~(\cite{DBLP:conf/ssbse/CallanP21}).

\subsection{Multi-Objective Optimization}

Performance properties such as runtime and memory consumption often are at odds with each other, i.e., one can improve runtime by caching results, thus increasing memory use, and vice versa. 
In order to improve such conflicting properties, multi-objective (MO) algorithms have been proposed~(\cite{DBLP:journals/ec/SrinivasD94}), which produce a Pareto front of non-dominated solutions. 
A solution $x$ Pareto dominates another $y$ if all of $x$'s objectives are as good as $y$'s and at least one objective is better than $y$'s.

Past work utilising MO algorithms for GI is sparse, with the majority of work focusing on single-objective improvement.
However, in the work where MO improvement has been successful Genetic Algorithm (GA) based algorithms have been used.
\cite{DBLP:conf/gecco/WuWHJK15} and \cite{DBLP:conf/ssbse/CallanP22} used NSGA-II~(\cite{DBLP:conf/ppsn/DebAPM00}), \cite{DBLP:journals/tec/WhiteAC11} used SPEA2~(\cite{DBLP:conf/ppsn/KimHMW04}), with \cite{DBLP:conf/kbse/MesecanB0CP22} comparing four MO algorithms, with SPEA2 and NSGA-III~(\cite{DBLP:journals/tec/DebJ14}) performing best.

In each algorithm, a population of solutions (in our case program variants) is generated and their fitnesses are measured.
In order to generate new patches, mutation, and crossover operators are used to generate child populations and then individuals are selected for the next generation from both child and parent populations.

The algorithms vary in their selection phases.
The algorithms use Pareto dominance  to compare different individuals who may find trade-offs between different properties.

%\begin{figure}
%    \centering
%    \includegraphics[scale=0.23]{NSGAII.png}
%    \caption{A diagram showing the selection phase of NSGA-II P is the parent population, and Q the offspring. They are split into Pareto fronts and the next generation is selected from the fronts in ascending order. If a front is too large for the next generation the front is sorted by a crowding metric and the least crowded portion of the list is taken to fill P'.
%    }
%    \label{fig:nsgaii}
%\end{figure}
%
%\begin{figure}
%    \centering
%    \includegraphics[scale=0.15]{NSGAIII.png}
%    \caption{A diagram showing the niching phase of NSGA-III. This problem has 3 reference lines, drawn from the origin (the ideal point) to 3 reference points, and 2 objectives.}
%    \label{fig:nsgaiii}
%\end{figure}
Both NSGA-II and NSGA-III sort the population into Pareto fronts based on their fitnesses. 
The population of the next generation is then selected from the top fronts, one at a time, until a set number of individuals are chosen.
If a front needs to be split, as it is too big for the population size, it is sorted by a crowding metric, and the least crowded members are selected.
%This is shown in Figure~\ref{fig:nsgaii}. 
In NSGA-II, crowding is based on distance from other individuals in the fitness landscape.
Whereas in NSGA-III, crowding is based on reference lines and the number of individuals that are closest to them, or niched to them.
%This is illustrated in Figure~\ref{fig:nsgaiii}.
NSGA-III selects individuals spread across as many niches as possible in the final front to maintain diversity. 

Unlike the NSGA algorithms, SPEA2 does not separate the population into Pareto fronts. 
Instead, the strength of each individual is calculated. 
This is equal to the number of other individuals that the individual Pareto dominates.
The raw fitness of an individual is then calculated as the sum of the strengths of all other individuals which it dominates. 
Like the NSGA algorithms crowding metric is calculated.
For this, all other individuals are sorted into a list based on proximity to the individual of interest.
The metric is inversely proportional to the distance of the $kth$ individual in the list.
The parameter $k$ is equal to the square root of the total population size.
Finally, the raw fitness and the crowding metric are simply added together and used to select individuals.

It is yet unclear which multi-objective approach works best for the purpose of genetic improvement, thus we explore the capabilities of these three algorithms shown successful in previous work.

\section{Multi-Objective GI for Android}
\label{sec:GI}
%We propose using multi-objective Genetic Improvement to optimize multiple different types of performance for Android Applications.
%We propose a framework which is specifically tailored to the improvement of Android apps, with novel mutation operators inspired by changes made by human developers. 

%\subsection{Challenges of Applying GI to Android}

%Although multi-objective search algorithms exist, that can help find trade-offs between various properties~\cite{DBLP:conf/ppsn/DebAPM00,DBLP:journals/tec/DebJ14,DBLP:conf/ppsn/KimHMW04}, they have thus far rarely been applied in the field of GI.

% Motivation
% We wish to improve multiple NFPs simultaneously, in a way that can be easily applied to any application, and any part of an application which displays performance issues.
% Thus, we choose to target the application's source code, all applications implement most of their functionality through their source code and source code can often be modified to improve NFPs.

% We wish to improve multiple properties at once as some properties may conflict, improving one property may come at the cost of another and if we do not account for this our patches may end lowering the overall quality of the application.

% Challenges

There are a number of practical changes when using genetic improvement to enhance the performance of Android apps when compared to traditional desktop environments.

Android applications make use of APIs for features like UI elements which are only present on actual devices.
The Android library available when testing applications on desktop operating systems overwrites the APIs such that they throw errors when invoked.
Most Android code utilises the Context class~(\cite{context}), in the applications we use in our experiments, the context class is explicitly imported in over 1/3 of files. 
This does not include the instances where it is implicitly imported as a nested dependency. 
This class gives the code access to the shared state of the application but is only available on devices.
This means that in order to run tests that exercise any component of an application's code that accesses this state, the entire application must be compiled, packaged, transpiled, installed, and launched before it can be tested.
This can take a considerable amount of time, often longer than the tests themselves~(\cite{DBLP:conf/ssbse/CallanP21}).

Android apps are generally built using Gradle with the Android Gradle plugin.
This makes them incompatible with much of the tooling surrounding automatic compilation and testing of code~(\cite{androidTest,androidBuild}).\looseness=-1

Another challenge of applying GI to the mobile domain is the accurate measurement of the fitness function.
Previous work has only applied GI to problems that take seconds/minutes to run. 
In the mobile domain, it was shown that apps that run in more than 150ms are considered to be `laggy' by users~(\cite{tolia2006quantifying}).
Therefore, although previous work used approximate fitness measurements, these are not appropriate in the mobile domain as they may not capture such minor, yet important, differences in non-functional behaviour.

% Framework

%\subsection{MO-GI for Mobile Apps}

\begin{figure}[t]
    \centering
    \includegraphics[scale=0.6]{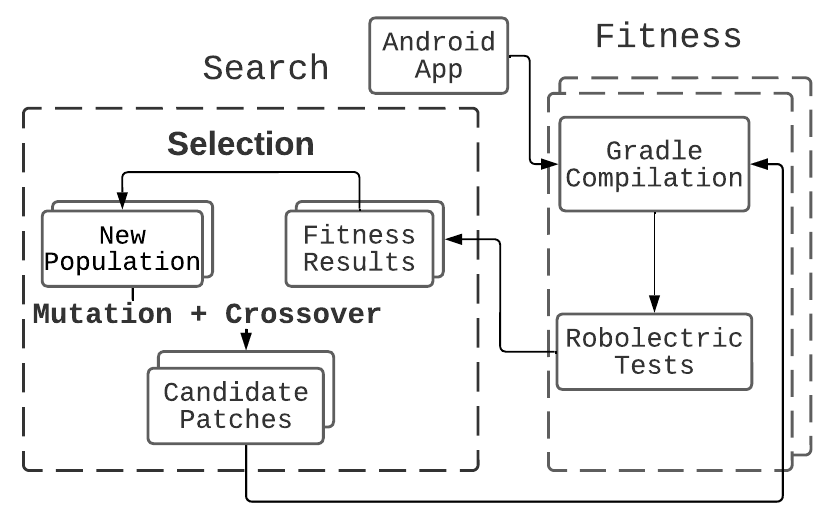}
    \caption{GI framework for Android app improvement, with search based on a genetic algorithm. In the case of local search, only mutation is applied.}
    \label{fig:framework}
\end{figure}

In order to overcome the aforementioned challenges, we propose the GI framework shown in Figure~\ref{fig:framework}.
The framework is split into two main components: the Search, and the Fitness sections.
These components can be swapped out depending on the properties being improved.

\subsection{Representation}
We use a program representation consisting of a list of edits, which are applied sequentially to the source code.
This representation has been used in GI many times in the past and proven successful~\cite{DBLP:journals/tec/PetkeHHLWW18}.
We use a list of edits, rather than representing the whole program in the genome, as may be done in traditional genetic programming, to reduce the memory footprint of the search process.
An example of this representation, as used in \tool, is shown in Figure~\ref{fig:patch}.

\begin{figure}
    \centering
    \begin{lstlisting}
        | gin.edit.statement.DeleteStatement Example.java:608 
        | gin.edit.statement.CopyStatement Example.java:1307 ->
        Example.java.java:320:365 |
    \end{lstlisting}
    \caption{An example of a program variant that deletes the statement with ID 608 and then copies the statement with ID 1307 to position 265 into the block with ID 365 in the file Example.java}
    \label{fig:patch}
\end{figure}

% Justyna: I know one of reviewers wanted it but it did not fit in here. Might move to the Background section if we get a request like that again.
%A patch $\delta$ is a sequence edits $e$ from the set of defined edits $E$ are that are applied to the target program $P$, resulting in variant $v$.
%This is illustrated in Equation~\ref{eq:mut}~(\cite{DBLP:journals/ese/PetkeABBWW23}).
%\begin{equation}
%    \label{eq:mut}
%    v = \delta(P) = e_k (e_{k-1}(...(e_1(P))), e_i \in E 
%\end{equation}
%By using different sequences of edits we can explore the search space $S$ of all potential variants.
%We can generate new variants using mutation and crossover operators on our sequence of edits.
%We can guide our exploration of $S$ with a fitness function, $f : S \rightarrow \mathbb{R}$, which can evaluate how good a variant is with respect to property of interest.\looseness=-1

\subsection{Fitness}

In the Fitness section in our framework (see Figure~\ref{fig:framework}), we measure the properties that we are improving.
As in previous GI work, we patch the application, compile it and run unit tests on it.
If all unit tests pass, the patch is considered valid, if not, it is discarded.
Then, the property being improved is measured. 
For example, if we are improving execution time, the time taken by the test suite is measured.
Multiple different properties are measured in the case of MO improvement.

Due to the complexity of the Android build system and significant use of UI elements, a minor change usually requires a time-costly process of installation on the actual device for testing.
Our framework thus utilises only the local tests which run on the JVM.
This would normally limit the components that could be tested to only those which do not use the device-only APIs.
If we attempt to use these APIs in a local test, we will simply call stubbed versions of the methods which throw exceptions.
However, by using the simulation-based Robolectric testing library, we are able to test any application component with fast local tests.
Robolectric has two main features that allow us to test apps.
Firstly, the simulation of the application and Android environment, which creates a headless version of the application within a local JVM.
Secondly, shadowing which allows the bytecode of classes to be overwritten at runtime. 
This is used to overwrite the API calls with simulated API calls and allows the simulated app to be exercised.
Shadowing is useful for mocking hard dependencies and can be used to avoid the complex setup needed when testing certain components.
Using this simulation-based approach, we can quickly compile and test application variants, and use measurement tools that aren't available in the Android operating system.
\cite{DBLP:conf/icse/CallanP22} found that improvements that could be demonstrated with unit tests written in the Robolectric library translated to improvements on Android applications run on real devices, in every case where improvements were found.
Thus, with a combination of Robolectric testing and manual review of improvements, we can be confident whether we have found an actual improvement or not.
% We use the Robolectric library to test code that utilises the Android-specific APIs which are not normally available on desktop devices.
We use the Gradle build system with the Android plugin to compile and test applications.

\cite{DBLP:journals/software/KhalidSNH15} identified execution time, memory, bandwidth, and energy usage as the most complained about and impactful non-functional properties of Android apps.  
In this work, we will attempt to improve execution time, memory, and bandwidth.
%Both execution time and memory usage have been successfully improved in the past with GI.
%JP: commented out as it;s not clear why we even mention it, and I don;t htink it;s needed, (depending on what sort of study you find about what users care about)
%We excluded frame rate as it is not possible to measure with local unit tests.
%One attempt has been made to improve the frame rate of Android applications in the past, using instrumented tests, however, failed to find any significant optimisations~\cite{DBLP:conf/ssbse/CallanP21}.
%We consider improving the energy usage of applications, however 
Previous work on automatically improving energy usage of Android apps~(\cite{DBLP:conf/gecco/BokhariBA017,DBLP:journals/tse/MoralesSKCA18}) found energy estimates to be too noisy, thus requiring external devices for physical energy measurements.
%For instance, Bokhari~et~al.~\cite{DBLP:conf/gecco/BokhariBA017} took great care to control the measurements, with use of external devices, and found that long configuration times and charging breaks were needed to achieve accurate measurements. 
Although GI can be used to optimize energy consumption~(\cite{DBLP:conf/gecco/BokhariBA017}), we want to provide a general, easy-to-use tool that does not require extra hardware.
%Previous work on improving the energy usage of Android applications found that fitness measurements were subject to noise from many sources, including mobile device temperature. 
%Bokhari~et~al.~\cite{DBLP:conf/gecco/BokhariBA017} took great care to control the measurements, with use of external devices and found that long configuration times and charging breaks were needed to achieve accurate measurements. 
%Similarly, Morales~et~al.~\cite{DBLP:journals/tse/MoralesSKCA18} use external measuring device.  
%Therefore, we 
It is worth mentioning that thus far the primary technique for improving 
bandwidth has been prefetching~(\cite{prefetchAds}).
No attempts have been made to improve it using source-code transformations, despite such changes being made by developers~(\cite{DBLP:journals/ese/CallanKPS22}).
We are the first to try to do so.

\subsection{Search}
The Search section of our framework for Android app improvement (see Figure~\ref{fig:framework}) determines how the search space of patches is navigated.
Most GI work so far has used single-objective algorithms, such as genetic programming and local search. 
Only a few consider more than one objective.
We pose that consideration of multiple objectives in the mobile domain is especially important, due to limited resources.
To fill this gap, we propose to utilise multi-objective approaches in the search process.
%Different search algorithms can be used here, depending on the number of properties being improved. 
%In the case of genetic algorithms, populations will be produced and improved by applying changes to existing patches.
%These are applied via mutation and/or crossover operations.
%Mutations typically delete, copy, or replace fragments of code, e.g., statements, while crossover appends two lists of changes together, though others have been tried~(\cite{DBLP:journals/tec/PetkeHHLWW18}).
Multi-objective algorithms will allow us to evolve patches that will balance different trade-offs, producing Pareto fronts of solutions.
The user will then be left with a choice of which patch fulfills their particular needs.
The multi-objective approach will provide relevant information on how runtime reductions might for impact memory use etc.

To start our search we need to generate an initial set of patches.
Our patch representation is not of fixed size and may contain any number of edits.
We create an initial population containing individuals consisting of single random edits.
Further creation is guided by a given search algorithm, where mutations and crossover are applied to create new patches.

\subsubsection{Mutation and Crossover}

Patches are created via mutation and crossover on the list of edits representation.
In the single-objective search used in GI so far crossover typically appends the lists of edits together from patches selected using binary tournament selection.
We apply this type of crossover in our MO algorithms as well.
%In our case, crossover operators, as implemented in the given multi-objective algorithms, are used.
A mutation simply adds or deletes an edit. 
In our case we operate on the statement-level, thus each mutation can delete, remove, or replace another statement.
Additionally, we investigated which other mutation operators might be beneficial in the Android domain.

%making GI an impractical approach~\cite{DBLP:conf/gecco/BokhariBA017}.
%Finally, we consider improving the bandwidth usage of applications.
%We propose using GI to find general source code transformations to improve bandwidth usage.
%, we chose to improve the execution time, memory consumption, and bandwidth of apps.

\cite{DBLP:journals/ese/CallanKPS22}.'s work showed that one of the most common techniques for improving non-functional properties of Android apps is caching.
Caching was found to be effective across all properties studied and improved a number of different applications in different domains.
Outside of the changes already implemented by standard GI mutation operators (remove code, change order of operations), caching is the most generically applicable strategy found, and thus, the most suitable for multi-objective improvement.
%This kind of change would not be possible using the standard mutation operators used in GI, thus we seek to introduce new operators which can make these kinds of changes.
Based on manual analysis of the commits from Callan et al.'s work, in which caching is used, we propose two new mutation operators.
Caching could prove useful for the three properties which we wish to improve.
Firstly, if we no longer need to execute a method as we already have the result we will save time.
If the method has a larger memory footprint than the stored result, we will reduce the memory footprint of the app.
Finally, if the cached method accesses the network, we will be able to avoid this operation and reduce network usage.
However, caching may negatively impact memory usage if the stored result is large.
This will mean that we will have to consider possible tensions between objectives when we run our search.

First, we propose a simple \textbf{In-Method Caching Operator}.
This operator simply stores the result of calling a method in a local variable and replaces future calls to this method with the local variable (see Algorithm~\ref{alg:method}).
An example of this operator can be seen in Figure~\ref{fig:inmethod}.
The second caching operator creates new fields in the associated class for storing cached method calls. 
This \textbf{Class Caching Operator} allows cached variables to persist beyond the end of individual method calls and could prove particularly useful if a method is called repeatedly. 
An example of this operator is shown in Figure~\ref{fig:classcache}.
We wrap the statement which accesses the cached variable with a null guard so that the first time it is called we actually call the method.
For both of these operators, we consider method call expressions to be cachable to the same variable only if their arguments consist of the same variables.
As shown in Algorithm~\ref{alg:class}, the class caching operator can be applied to any method call expression.
However, as local variables do not persist after a method is executed, there must be at least two instances of the expression for it to be cached.
These operators will not disrupt the source code syntax as they simply replace a method call expression with a variable name expression which is the same type as the method's return type.

\begin{algorithm}[t]
		\caption{Find method calls to cache in Method M}
		\label{alg:method}
		\begin{algorithmic}[1]
			\Function{MethodCacheFinder}{$C$}
			\State $seen = \emptyset$
                \State $cachable = \emptyset$
			\For {each expression $e$ in $M$}
			\If{$e$ is a method call expression}
                \If{$e \in seen$}
                \State $cachable = cachable \cup e$
                \Else
                \State $seen = seen \cup e$              
                \EndIf
                \EndIf
			\EndFor
			\State Return $cachable$
			\EndFunction
		\end{algorithmic}
\end{algorithm}

\begin{figure}[t]
    \centering
    \begin{lstlisting}[escapechar=@,basicstyle=\small]
@\textbf{Original Code}@           @\textbf{Mutated Code}@

int x = foo(a,b,c);  @\textbf{int cachedVar1 = foo(a,b,c);}@
int y = foo(a,b,c);  int x = @\textbf{cachedVar1}@;
                     int y = @\textbf{cachedVar1}@;
    \end{lstlisting}
    \caption{An example of the In-Method Cache Operator. The resultant code stores the results of a method call $foo$, with parameters $a$, $b$ and $c$. This stored result can then be used later in the same method.}
    \label{fig:inmethod}
\end{figure}

\begin{algorithm}[t]
		\caption{Find method calls to cache in Class C}
		\label{alg:class}
		\begin{algorithmic}[1]
			\Function{ClassCacheFinder}{$M$}
                \State $cachable = \emptyset$
                \For{each method $m$ in $C$}
			\For {each expression $e$ in $m$}
			\If{$e$ is a method call expression}
                \State $cachable = cachable \cup e$
                \EndIf
                \EndFor
			\EndFor
			\State Return $cachable$
			\EndFunction
		\end{algorithmic}
\end{algorithm}

\begin{figure}[t]
    \centering
    \begin{lstlisting}[escapechar=@, basicstyle=\small]
@\textbf{Original Code}@           @\textbf{Mutated Code}@  

class C1 {           class C1 {
public void foo(){   @\textbf{int cachedVar1;}@
    int x = a();     public void foo(){ 
    }                  @\textbf{if (cachedVar1 == null)\{}@
}                        @\textbf{cachedVar1 = a();}@
                       @\textbf{\}}@
                       int x = @\textbf{cachedVar1;}@
                       }  
		     }
    \end{lstlisting}
    \caption{An example of the Class Cache Operator. The result of a method call is stored in a field of the class for later use in any method.}
    \label{fig:classcache}
\end{figure}

\section{Research Questions}

\label{sec:RQs}

To evaluate how effective the multi-objective GI approach for improvement of Android apps' runtime, memory, and bandwidth use is, we pose the following research questions:

%\begin{itemize}
\smallskip
%    \item [RQ1] 
\noindent \textbf{RQ1: Can MO-GI optimize Android apps in the same way as real developers?}
    
\noindent In order to validate our approach, we want to see if MO-GI can reproduce real-world improvements that Android developers have manually implemented in the past.

\smallskip
%    \item [RQ2] 
\noindent \textbf{RQ2: How effective is MO-GI at optimising Android apps without known improvements?}
    
\noindent Answering this question will allow us to see how well our approach generalises. In particular, if it's able to find improvements in current code.
%We show how developers could use our approach to find new improvements.

\smallskip
%    \item[RQ3] 
\noindent \textbf{RQ3: Which MO algorithm is the most effective for MO-GI for Android?}

\noindent There are a number of different MO algorithms available. We want to ensure that our approach utilises the most effective one, thus we investigate and compare a selection of MO algorithms successfully used in the GI domain in the past.

\smallskip
%    \item [RQ4] 
\noindent \textbf{RQ4: How do the improvements found by MO-GI compare to those found by SO-GI for Android apps?}

\noindent    We wish to see if using MO algorithms limits GI's ability to improve apps, when compared to improving only a single objective.
This is especially important in cases where one improvement can enhance two objectives (e.g., deletion can improve both runtime and memory use). 
We want to see if MO are still competitive in such cases.

\smallskip
%    \item[RQ5] 
\noindent \textbf{RQ5: What is the runtime cost of MO-GI for Android?} 

\noindent Any improvements found by MO-GI must be considered against the cost of running it.
    The improvements found must justify this cost. 

\smallskip
%    \item[RQ6] 
\noindent \textbf{RQ6: How does GI compare with available state-of-the-art techniques for Android performance improvement via code modification?}

\noindent We want to compare \tool~with state-of-the-art tools that are readily available to developers to see if our tool could offer an attractive alternative.
%\end{itemize}

\section{Methodology}
\label{sec:Meth}

In order to answer our research questions, we propose a series of experiments, running both multi- and single-objective GI on a benchmark of real-world Android applications.

To answer \textbf{RQ1}, \textbf{RQ3}, and \textbf{RQ5}, we run GI with three multi-objective algorithms on a set of applications, in some of which we know potential improvements are present, in order to validate our approach.
To answer \textbf{RQ2}, we use the same setup to improve the latest versions of applications, to see if our framework can find yet-undiscovered optimizations 

Next, to answer \textbf{RQ4}, we run GI with a single-objective hill climbing algorithm, to compare with a multi-objective approach.
With this set of experiments, we can evaluate whether or not our multi-objective algorithms can find improvements that are as good as those found by single-objective search.
This allows us to compare the trade-offs found by different search algorithms.

Finally, to answer \textbf{RQ6}, we use an Android linter to identify performance issues within our benchmarks.
Linters are the only tools available to Android developers which can identify issues with source code that may affect performance properties we target.
    By manually repairing these issues we can see how our tool compares in terms of both effort and effectiveness with respect to existing tools available to developers.
%We then manually repair the issues that the linter highlights.
%This allows us to compare our approach to the 

\subsection{Genetic Improvement Framework}

We implement our multi-objective GI approach for Android in a tool called~\tool, and use it to answer our RQs. 
Although there are many existing GI frameworks, \cite{DBLP:conf/gecco/ZuoBP22} found that PYGGI~(\cite{An:2019:PLI:3338906.3341184}) and the Genetic Improvement In No time tool (Gin)~(\cite{DBLP:conf/gecco/BrownleePABWW19}) were the only GI tools that could be readily applied to new software, with more recent tool by \cite{DBLP:conf/kbse/MesecanB0CP22} not yet available.
However, none of the aforementioned work can be run upon Android applications.
Whilst Gin is compatible with most Java programs, and thus could potentially easiest to extend, it is not compatible with the Android compilation and testing environments.

In \tool{}, we implement three MO algorithms:
NSGA-II~(\cite{DBLP:conf/ppsn/DebAPM00}) as it is one of the most widely used multi-objective algorithms;
NSGA-III~(\cite{DBLP:journals/tec/DebJ14}), that was specifically developed for problems with 3 or more objectives in mind;
and SPEA2~(\cite{DBLP:conf/ppsn/KimHMW04}), which has recently proven successful for MO-GI in the desktop domain~(\cite{DBLP:conf/kbse/MesecanB0CP22}).
We use MO algorithms, as we believe that we will be able to find better improvements to some properties if we are able to sacrifice others.
In particular, with our caching operators -- these operators are likely to negatively impact the memory consumption of the applications, however a small increase in memory consumption may be worth it if it can sufficiently improve another property.
The parameters used in these implementations can be found in Table~\ref{tab:params}. 

To measure execution time we use Linux's time tool~(\cite{LinuxTime}), we measure memory usage with the Java Runtime's memory allocation tracking~(\cite{JavaRuntime}) and we use Linux's built-in process-level network traffic tracking~(\cite{LinuxProc}) to measure bandwidth.

\subsection{Benchmarks}

Genetic improvement requires a set of tests that cover the areas of code being improved, in order to validate that a non-functional property-improving patch does not negatively affect the app's functionality.
Unfortunately, most open-source Android applications do not have test suites, and those that do are limited, achieving a median line coverage of 23\%~(\cite{DBLP:conf/iwpc/PecorelliCFLP20}).
Furthermore, there is not a single tool that we have found in an extensive search of the literature which can automatically generate unit tests for Android applications.
All automated testing tools for Android found~(\cite{DBLP:conf/ssbse/AuerAF22,DBLP:conf/sigsoft/AmalfitanoAFT15,DBLP:conf/oopsla/AzimN13,DBLP:conf/kbse/BaekB16,DBLP:conf/sigsoft/MahmoodMM14,DBLP:conf/issta/MaoHJ16,DBLP:conf/sigsoft/SuMCWYYPLS17,DBLP:conf/icse/LiYGC17,DBLP:journals/symmetry/YasinHY21}) focus on testing UI via input generation in order to induce crashes and only run on devices/emulators, so would not be compatible with our framework.
Moreover, they do not generate assertions --- crucial for capturing correct app behaviour.

This meant that we had to manually construct unit tests for every single benchmark.
We first had to attempt to understand each application and the component being improved and then attempt to create thorough, high-quality tests for them.
In many cases, we had to account for asynchronous code, which was scheduled by the target code, and ensure that it executed completely during test execution.
In other instances, we had to hunt down various parts of the state of the application to ensure they were correct.
%As previously described, in many cases tests require APIs that are not available to local JUnit tests running on a desktop operating system.
%Alternatively, the Robolectric testing library allows code that would normally have to be run on an Android device or emulator to be run locally and faster.
%We, therefore, chose to write tests using the Robolectric library and we used the await library to allow synchronous testing of asynchronous components.
For each test we created, we ensured that it covered the methods which we wished to improve.
We also added assertions about the state of the components of the application that were modified during execution.
We achieved at least 75\% branch coverage for methods used in our study.
%This process would take hours and in some cases whole days depending on the complexity of the code being tested.
We do, however, note that developers would find this process simpler, as they already have an understanding of the application.
They would get many other benefits from writing tests~(\cite{DBLP:conf/esem/MockusND09, DBLP:conf/esem/BachAPL17})
so the cost cannot be only placed upon the application of GI.
Given the cost associated with manual testing, we set a threshold of 20 benchmarks for all our experiments.

To validate our approach, we first run \tool~on applications with known performance issues.
\cite{DBLP:journals/ese/CallanKPS22} has recently conducted a study of the changes that Android developers make to improve app performance.
They pose that some of those changes are within the GI search-space.
For instance, moving an operation outside of a {\sc for} loop, if only need to be executed once.
While others are not yet achievable, e.g., requiring new code to be added that could not be achieved via mutation of the existing code base.
We thus use \cite{DBLP:journals/ese/CallanKPS22}'s criteria to iteratively analyse the commits from their dataset that improve runtime, bandwidth, or memory use, until we reached our 20 benchmark target.
In particular, we found 14 commits in previous work, spread over different versions of 7 applications.
Since we also want to find improvements in current software, we stop our selection procedure here and use the current versions of the 7 apps, giving us a total of 21 benchmarks.

Once we had our set of versions of apps, we prepared them for GI.
Firstly, we had to ensure the apps would build.
Over time, a number of changes have been made to the Android build tools, making older versions of code incompatible with modern Android Studio.
We require these build tools to function with Android Studio, so we can test and measure the test coverage of applications confirming that they can be safely improved.
This meant that we had to update build scripts with newer versions of libraries and build tools.
In some cases, there were bugs such as unescaped apostrophes in resource files, which prevented applications from building.
These bugs were fixed.
In a few cases, the benchmarks also used outdated non-Gradle build systems, so we wrote the necessary build scripts, and modified the project's directory structure, to be compatible with Gradle and thus with \tool. 
No source code was modified in this process.

We ran the PMD static analyser on the 7 applications and ran \tool~on the classes which showed the most performance issues. 
This way we could see how our approach compares against human effort for finding performance-improving code transformations of existing code bases, for the 14 previously patched app variants.
We could also see whether our approach is able to find yet unknown performance improvements in the current versions of the 7 apps.
%This resulted in 21 benchmarks, 7 current versions, and 14 previous.

% \begin{table}
%     \centering
    
%     \caption{Commits which improve Android applications which GI could reproduce.}
%     \begin{tabular}{ll}
%         \toprule
%         Application & Improving Commits\\
%         \midrule
%         \multirow{5}{*}{PortAuthority} & e0163e20d1a67c22c2f7ed0f0345206ce1a050f0\\
%         & e37a1a522a15773710f051d9fff5c0ce68ade5cb\\
%         & 3a1329297881aff069cdbc80c92de386ac952d77\\
%         & 3e6846b6a377c35780ddb49e21eeab5749381bf2\\
%         & a02a0170a38ec257e1f390388e4b5d1414b3cf36\\
%         \midrule
%         \multirow{2}{*}{Tower Collector} & 956ea2213c1f7f012d6ab1388536a0c6d5202bd9 \\
%         & 0632608d26667e3a1864bf436086cf9422a913cb\\
%         \midrule
%         \multirow{1}{*}{Gadgetbridge} & c75362c5ea489247cc00b473a0ef91dbb1cc1569 \\
%         \midrule
%         Fosdem Companion &  b79e29a67c29699b9b8d4ad9c09a3349ce32c59f \\
%         \midrule
%         \multirow{2}{*}{Fdroid} & e44ca193dd0adcbc5e240410aec4c681f5053dae \\
%         & bf8aa30a576144524e83731a1bad20a1dab3f1bc \\
%         \midrule
%         \multirow{1}{*}{Lightning Browser} &  460da386ec10cb82b97bd2def2724fe41f709a88 \\
%         \midrule
%         Frozen Bubble & e9f6a51be9f7c4ad9f11d8712b06cb906e9ddf28 \\
%         \midrule
%     \end{tabular}
%     \label{tab:commits}
% \end{table}

\subsection{Experimental Setup}
\begin{table}[]
    \centering
    \caption{Parameter settings for the MO algorithms used in our study.}
%Reference points are measured with respect to each improvement property and each benchmark.}
    \begin{tabular}{ll}
        \hline
        Parameter & Value \\
        \hline
        Mutation Rate & 0.5 \\ 
        Crossover Rate & 0.2\\ 
        No. Generations & 10\\ 
        No. Individuals & 40\\
        Selection & Binary Tournament \\
        Crossover & Append Lists of Edits \\
        Mutation & Add/Remove an Edit \\
        Reference Points & Worst Observation (for \\
	& each prop. and bench.)\\
        \hline

    \end{tabular}
    \label{tab:params}
\end{table}
For each version of code we improve, we run \tool{} 20 times with 400 evaluations.
To minimise measurement noise, we use the Mann-Whitney U test at the 5\% confidence level to determine whether there is an improvement of a given property (i.e., runtime, memory or bandwidth use).
For the evolutionary algorithms, we divide these 400 evaluations into 10 generations with 40 individuals each, as  was shown to be effective in previous work, including in the Android domain~(\cite{DBLP:journals/tse/MotwaniSBJG22, DBLP:conf/icse/CallanP22}).
We set number of evaluations to 400 as, even when using simulation-based testing, the evaluation of an individual is slow, taking up to 2 minutes.
We use the Genetic programming parameters in Table~\ref{tab:params} as they have been used successfully in the past~(\cite{DBLP:conf/ssbse/CallanP22}).

We had 2520 runs in total, taking a mean of 3 hours per run, resulting in roughly 7500 hours of computing time to test our approach.

All of our experiments were performed on a high-performance cloud computer, with 16GB RAM and 8-core Intel Xenon CPUs.
We ran jobs across 10 nodes, each running separately to avoid interference between fitness measurements. 

\section{Results and Discussion}
\label{sec:Res}

In this section, we present and analyse the results of our experiments, answering our Research Questions (Section~\ref{sec:RQs}).
Throughout this section we will refer to the CPU time (s) of the test process as execution time, the size of the occupied Java heap as memory consumption (MB), and the number of bytes sent and received by the test process as network usage (B).
Each of these objectives is a fitness function which we aim to minimize.
%We will first discuss how well \tool{} performs on the benchmarks in which developers found improvements in the past, based on commit logs to validate our approach.

\subsection{RQ1: Known Improvements}
\begin{figure}[t]
    \centering
    \includegraphics[scale=0.45]{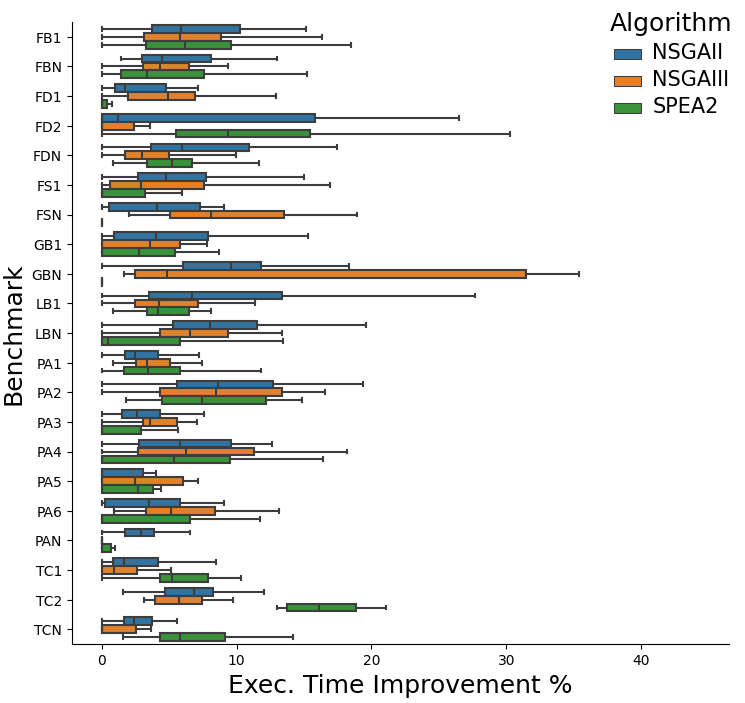}
    \caption{Execution time improvements (\%) achieved by \tool~using three MO algorithms on 21 versions of 7 Android apps.}
    \label{fig:exec}
\end{figure}

\begin{figure}[t]
    \centering
    \includegraphics[scale=0.45]{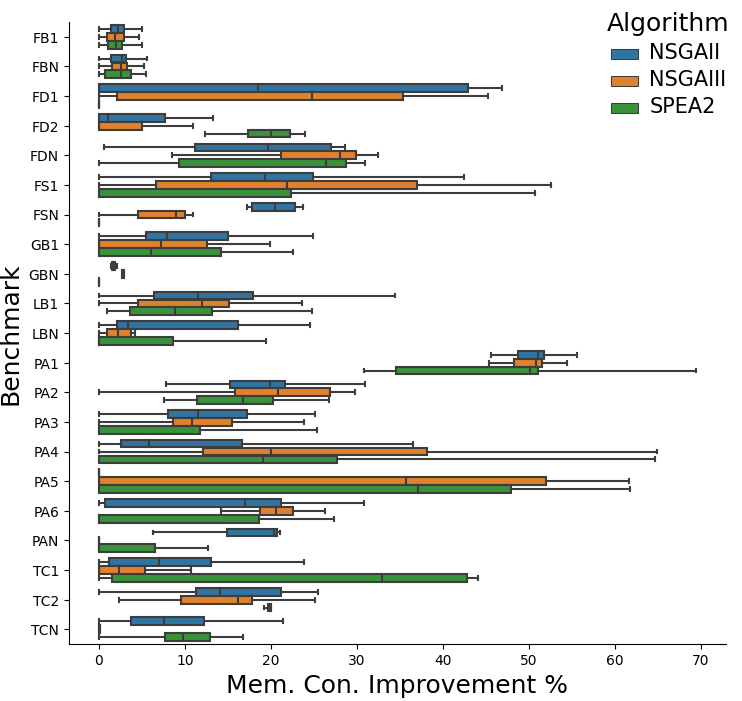}
    \caption{Memory consumption improvements (\%) achieved by \tool{} using three MO algorithms on 21 versions of 7 Android apps.}
    \label{fig:mem}
\end{figure}

Figure~\ref{fig:exec} and ~\ref{fig:mem} show the improvements found in the benchmarks in which we knew improvements were possible. 
We find improvements to both execution time and memory, but not bandwidth.
We believe this is due to the nature of the benchmarks.
Although feasible, only one application had bandwidth improvements in its history that would be achievable by GI.
This improvement\footnote{\url{https://github.com/erikusaj/fdroidTvClient/commit/bf8aa30a576144524e83731a1bad20a1dab3f1bc}}
 required 2 insertions and 2 deletions at once to be achieved and thus was more difficult to evolve over time.

We find improvements to execution time of up to 26\% and memory of up to 69\%.
We manually analysed the patches found in order to determine whether GI was capable of finding the same patches that developers made to improve their applications.
The result of this analysis can be found in Table~\ref{tab:knownImpMatch}. In ~64\% of benchmarks \tool~is able to find patches containing edits semantically-equivalent to developer patches, providing at least the same \% performance improvement.
In other words, aside from reproducing improvements, in some cases, we find additional edits, further improving app performance.

\begin{table}[t]
    \centering
    
    \caption{No. of times \tool{} finds patches that contain edits semantically-equivalent to developer patches, providing at least the same \% performance improvement (Rep.) and no. runs where an improvement was found (Imp.). Each MO run was repeated 20 times.}
    \label{tab:knownImpMatch}
    \begin{tabular}{lrrrrrr}
        \toprule
        Application Version & \multicolumn{2}{r}{NSGAII} & \multicolumn{2}{r}{NSGAIII} & \multicolumn{2}{r}{SPEA2}\\
        & Rep. & Imp. & Rep.& Imp. & Rep. & Imp. \\
        \midrule
        Port Authority 1 & 4 & 16 & 8 & 18 & 3 & 15\\
        Port Authority 2 & 0 & 17 & 0 & 15 & 0 & 14\\
        Port Authority 3 & 0 & 13 & 0 & 14 & 0 & 18\\
        Port Authority 4 & 4 & 15 & 8 & 17 & 10 & 13 \\
        Port Authority 5 & 5 & 19 & 3 & 19 & 0 & 12\\
        Port Authority 6 & 4 & 13 & 7 & 18 & 2 & 11\\
        \midrule
        Tower Collector 1 & 10 & 14 & 6 & 13 & 8 & 20\\
        Tower Collector 2 & 0 & 15 & 0 & 18 & 0  & 19\\
        \midrule
        Gadgetbridge 1 & 0 & 15 & 0 & 12 & 0 & 13\\
        \midrule
        Fosdem Companion 1  & 3 & 13 & 4 & 12 & 7 & 14  \\
        \midrule
        Fdroid  1 & 0 & 19 & 0 & 17 & 0 & 13\\
        Fdroid 2 & 8 & 14 & 4 & 12 & 6  & 16\\
        \midrule
        Lightning Browser 1 & 2 & 12 & 3 & 18 & 4 & 17 \\
        \midrule
        Frozen Bubble 1 & 13 & 15 & 12 & 16 & 12 & 18\\
        \midrule
    \end{tabular}
\end{table}
\begin{table}[t]
    \centering
    \caption{Normalised Hypervolumes of the Pareto fronts found by \tool{} across our experiments, by algorithm.}
    \label{tab:Hyper}
    \begin{tabular}{lrrr}
        \toprule
        Application Version & NSGAII & NSGAIII & SPEA2\\
        \midrule
        PortAuthority 1 (PA1) & 0.145 & 0.186 & \textbf{0.458} \\
        PortAuthority 2 (PA2) & 0.223 & 0.267 & \textbf{0.327} \\     
        PortAuthority 3 (PA3)& 0.259 & \textbf{0.285} & 0.249 \\
        PortAuthority 4 (PA4)& \textbf{0.429} & 0.225 & 0.112 \\
        PortAuthority 5 (PA5)& \textbf{0.247} & 0.073 & 0.196 \\
        PortAuthority 6 (PA6) & 0.053 & 0.053 & \textbf{0.143} \\
        PortAuthority Current (PAN)  & 0.051 & 0.133 & \textbf{0.59} \\ 
        \midrule
        Tower Collector 1 (TC1)& 0.03 & 0.019 & \textbf{0.127 }\\
        Tower Collector 2 (TC2)& 0.027 & 0.052 & \textbf{0.088} \\
        Tower Collector Current (TCN) & 0.254 & 0.017 & \textbf{0.309} \\
        \midrule
        Gadgetbridge 1 (GB1) & \textbf{0.611} & 0.568 & 0.158 \\ 
        Gadgetbridge Current (GBN) & 0.008 & \textbf{0.384} & 0.007 \\ 
        \midrule
        Fosdem Companion 1 (FS1)& 0.318 & \textbf{0.383} & 0.359 \\ 
        Fosdem Companion Curr. (FSN) & 0.105 & \textbf{0.138 }& 0.021 \\
        \midrule
        Fdroid 1 (FD1) & 0.016 & \textbf{0.206} &0.012 \\
        Fdroid 2 (FD2)  & 0.022 & 0.042 & \textbf{0.525} \\ 
        Fdroid Current (FDN) & 0.206 & 0.065 & \textbf{0.233} \\ 
        \midrule
        Lightning Browser 1 (LB1)& \textbf{0.322} & 0.159 & 0.028 \\ 
        Lightning Browser Curr. (LBN) & 0.038 & 0.037 & \textbf{0.039} \\
        \midrule
        Frozen Bubble 1 (FB1) & \textbf{0.097} & 0.094 & 0.076\\
        Frozen Bubble Current (FBN) & 0.024 & 0.024 & \textbf{0.026} \\
        \midrule
    \end{tabular}
\end{table}

% \begin{sidewaystable}
%     \centering
%     \begin{tabular}{lrrrrrrrrr}
%         \toprule
%         Application & \multicolumn{2}{r}{NSGA II} & \multicolumn{2}{r}{NSGA III} & \multicolumn{2}{r}{SPEA2}\\
%         \midrule
%         &Exec. Time & Memory Con. & Bandwidth &Exec. Time & Memory Con. & Bandwidth &Exec. Time & Memory Con. & Bandwidth  \\ 
%         \midrule
%         PortAuthority &8.9 & 23.1 & 0.0 &8.3 & 25.4 & 0.0 &7.8 & 28.8 & 0.0 \\
%         \midrule
%         Tower Collector  &16.7 & 21.5 & 0.0 &9.3 & 4.9 & 0.0 &20.6 & 16.8 & 0.0 \\
%         \midrule
%         Gadgetbridge &18.3 & 2.1 & 0.0 &35.4 & 5.3 & 0.0 &20.3 & 4.2 & 0.0\\
%         \midrule
%         Fosdem Companion &9.1 & 23.7 & 0.0 &18.9 & 11.0 & 0.0 &21.6 & 23.3 & 0.0 \\
%         \midrule
%         Fdroid &17.5 & 28.6 & 0.0 &26.0 & 32.5 & 0.0 &19.7 & 30.9 & 0.0 \\
%         \midrule
%         Lightning Browser &19.6 & 24.6 & 0.0 &14.3 & 11.6 & 0.0 &17.8 & 14.3 & 0.0 \\ 
%         \midrule
%         Frozen Bubble  &13.0 & 5.6 & 0.0 &15.8 & 5.3 & 0.0 &15.2 & 5.5 & 0.0\\
%         \midrule
%     \end{tabular}
%     \caption{Improvements on new versions of software}
%     \label{tab:newImps}
% \end{sidewaystable}

\subsection{RQ2: Improvements of Current Apps}
Next, we analyse the results of the experiments on the benchmarks of current versions of applications, to see how well our approach generalizes to code in which there are no known improvements.

The performance of each algorithm on versions of software is shown in Figure~\ref{fig:exec} and Figure~\ref{fig:mem}.
We find improvements to the execution time of up to 35\% and to memory consumption of up to 32\%.
Again no improvements were found to bandwidth.
We believe this is due to the nature of our benchmarks, where only FDroid 2 uses bandwidth extensively.

We have compiled the best changes found by \tool~in these experiments to demonstrate the capabilities of \tool.
%\textit{\textbf{RQ2.} \tool{} can improve execution time by \textbf{35\%} and memory usage by up to \textbf{32\%} on software without known improvements.}
% To validate the changes found by our method, we manually analyzed the individuals in the  Pareto front produced by our experiments.
% We then made pull requests\footnote{In the case of Fosdem-Companion, the whole application was translated into Kotlin, so we ran our experiments on the most recent Java version. But, were unable to make an actual pull request} with the best changes that we found with our experiments.
We detail each of these patches below:
\footnote{These can be found in our repository~(\cite{repo}) under `bestPatch' in each of the applications in the `Benchmark' folder.}
%\url{https://github.com/mogindroid/GIDroid/tree/main/Benchmark/PAN/bestPatch}} 

\subsubsection{Port-Authority (PAN)}

In the Port Authority application, our best change found
%\footnote{\url{https://github.com/mogindroid/GIDroid/tree/main/Benchmark/PAN/bestPatch}} 
consisted of removing an unnecessary try-catch statement, which resolved the IP address of a URL.
It would not only attempt to resolve URLs, but also, redundantly, IP addresses.
Furthermore, the resolved IP address is then passed to the constructor of the InetSocketAdress class, which already performs IP resolution, rendering the statement completely redundant.
The error handling is also performed in the same way when the IP address is passed to the InetSocket. 

\subsubsection{F-Droid (FDN)}
The improvement for F-droid
%\footnote{\url{https://github.com/mogindroid/GIDroid/tree/main/Benchmark/FDN/bestPatch}}
refactored an if/else statement.
Before, the statement checked if an object was null or not, instantiating it if it were null, and canceling its animation if not.
However, after this statement, the object was instantly re-initialised.
Meaning that in the case where the object was null, it would be instantiated once and then instantiated immediately after.
We refactor the statement to remove the null clause and only cancel the animation if the object is not null.
\subsubsection{Tower Collector (TCN)} 
In the TowerCollector, the best-evolved change
%\footnote{\url{https://github.com/mogindroid/GIDroid/tree/main/Benchmark/TCN/bestPatch}} 
 consisted of changes to the way in which a database is handled. 
It ensured that the connection to the database is closed when no longer needed and that the database is only instantiated when it is actually needed.
This change reduces memory usage but slightly increases execution time due to an extra function call.

\subsubsection{Frozen Bubble (FBN)}
In the Frozen Bubble application, the best improving change consisted of modifying how new rows of bubbles were instantiated in a row.
%\footnote{\url{https://github.com/mogindroid/GIDroid/tree/main/Benchmark/FBN/bestPatch}}
We found that checking for -1 in the newly generated row was redundant as the row cannot contain a -1, it can only contain positive integers.
We also found that the game pushed the sprite to the back of the board, but inspecting the application with and without this change shows no noticeable difference.
\subsubsection{Fosdem-Companion (FSN)}
In the Fosdem application the most improving change
%\footnote{\url{https://github.com/mogindroid/GIDroid/tree/main/Benchmark/FSN/bestPatch}}
 consists of moving the instantiation of two objects outside of a loop.
This means the same object can be reused in the loop, with the need for a new object to be assigned, thus saving both memory and execution time.
\subsubsection{Gadget-Bridge (GBN)}

In the best change for the GadgetBridge Application
%\footnote{\url{https://github.com/mogindroid/GIDroid/tree/main/Benchmark/GBN/bestPatch}}
 we cache the method call which resolves the name of a file that is repeatedly used and removes the redundant rendering of a view that is already visible.
\subsubsection{Lightning Browser (LBN)}
In Lightning Browser, the best-evolved mutation consists of removing a check for whether or not a list of bookmarks is null.
%\footnote{\url{https://github.com/mogindroid/GIDroid/tree/main/Benchmark/LBN/bestPatch}}
The list is an argument decorated with @NonNull so should never be null, and in the case that is is there will be no errors.

\subsection{RQ3: Multi-Objective Search}
\begin{figure}
    \centering
    \includegraphics[scale=0.6]{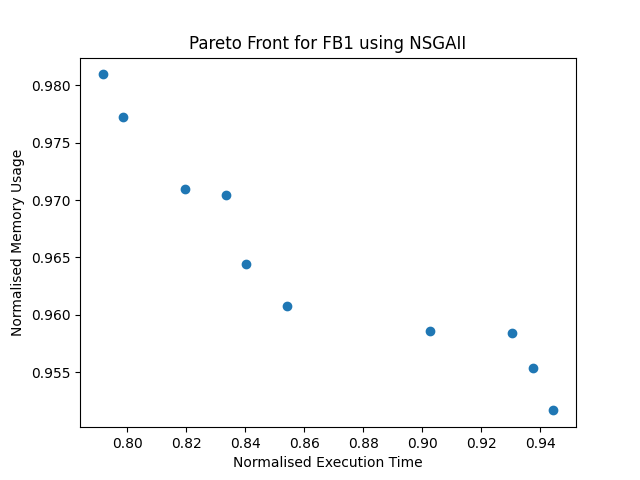}
    \caption{Pareto Front from NSGA-II experiments on the FB1 Benchmark.}
    \label{fig:front}
\end{figure}

In order to compare the different algorithms used in search, we consider the procedure proposed by \cite{LiTSE2020}, for comparing different multi-objective algorithms in a search-based software engineering context.
We choose to measure the hypervolume (HV) of the data, as it is considered to be a good indication of the general quality of the Pareto fronts produced and is considered appropriate when there is no preference between the different properties being improved.
In order to measure the hypervolume we specify the reference points as the worst observation for all fitness measurements, for each objective, as done in previous work~(\cite{DBLP:conf/icst/JiLCPZ0YL18,DBLP:journals/tcyb/LiuZL21}).
Due to different fitness scales, we normalise the values, though also present raw ones in our online repository, including all Pareto fronts~(\cite{repo}).
Normalised hypervolume values are presented in Table~\ref{tab:Hyper}.
The Pareto fronts from all of our multi-objective experiments can be found in our repository~(\cite{repo}).
We find that across our experiment we find patches spread across the Pareto front (see Figure~\ref{fig:front}), showing that trade-offs between properties must be considered in the search process, due to the natural tension between them.

\begin{table}[]
    \centering
    \caption{A effect size for each algorithm on each benchmark. Effect sizes larger than 0.5 show positive improvement. differences: N=negligible, S=small, M=medium, L=large}
\begin{tabular}{lrrrrrr}
    \hline
    \multirow{2}{*}{Benchmark} & \multicolumn{3}{c}{Exec. Time} & \multicolumn{3}{c}{Mem. Con.}  \\
    & NSGA-II & NSGA-III & SPEA2 & NSGA-II & NSGA-III & SPEA2 \\
    \hline
    PortAuthority 1 & 1.0 (L) & 1.0 (L) & 0.93 (L)& 1.0 (L) & 1.0 (L) & 1.0 (L)  \\
        PortAuthority 2 & 0.98 (L) & 1.0 (L) & 1.0 (L)& 1.0 (L) & 1.0 (L) & 1.0 (L)\\
        PortAuthority 3 & 0.97 (L) & 0.97 (L) & 0.97 (L)&  1.0 (L) & 1.0 (L) & 0.93 (L) \\
        PortAuthority 4 & 0.99 (L) & 0.99 (L) & 1.0 (L)& 1.0 (L) & 1.0 (L) & 1.0 (L) \\
        PortAuthority 5 & 0.67 (M) & 0.81 (L) & 0.18 (L)& 0.82 (L) & 1.0 (L) & 0.79 (M)  \\
        PortAuthority 6 & 0.88 (L) & 0.99 (L) & 0.71 (M)&  0.91 (L) & 1.0 (L) & 0.9 (L)\\
        PortAuthority Current & 1.0 (L) & 1.0 (L) & 0.67 (M) & 1.0 (L) & 1.0 (L) & 1.0 (L)\\ 
        \midrule
        Tower Collector 1 & 1.0 (L) & 1.0 (L) & 0.89 (L)& 0.98 (L) & 1.0 (L) & 0.92 (L)\\
        Tower Collector 2 & 1.0 (L) & 1.0 (L) & 1.0 (L)& 1.0 (L) & 1.0 (L) & 1.0 (L)\\
        Tower Collector Current &  0.92 (L) & 1.0 (L) & 0.85 (L) & 1.0 (L) & 0.67 (M) & 0.98 (L)\\
        \midrule
        Gadgetbridge 1 &  0.87 (L) & 0.96 (L) & 0.53 (N)&  1.0 (L) & 1.0 (L) & 0.54 (N)  \\
        Gadgetbridge Current 1 & 1.0 (L) & 1.0 (L)& 1.0 (L) & 1.0 (L) & 1.0 (L)& 1.0 (L)\\
        \midrule
        FosdemComp. 1 & 1.0 (L) & 0.95 (L) & 0.67(M) & 1.0 (L) & 1.0 (L) & 0.83 (L)\\
        FosdemComp. Current & 1.0 (L) & 0.95 (L) & 0.67(M)& 1.0 (L) & 0.83 (L) & 1.0 (L)\\
        \midrule
        Fdroid 1 & 0.77 (L) & 0.92 (L) & 0.73 (L) & 0.82 (L) & 1.0 (L) & 0.76 (L)\\
        Fdroid 2 & 0.99 (L) & 0.93 (L) & 0.92 (L) & 1.0 (L) & 1.0 (L) & 1.0 (L)\\
        Fdroid Current & 0.74 (L) & 1.0 (L) & 0.99 (L) & 0.98 (L) & 1.0 (L) & 0.99 (L)\\
        \midrule
        LightningBro. & 0.79 (L) & 1.0 (L) & 1.0 (L) & 1.0 (L) & 0.95 (L) & 1.0 (L)\\
        LightningBro. Current & 0.9 (L) & 0.83 (L) & 0.59 (S) & 1.0 (L) & 0.9 (L) & 0.92 (L)\\
        \midrule
        FrozenBubble 1 & 0.98 (L) & 1.0 (L) & 0.97 (L) & 0.98 (L) & 1.0 (L) & 0.97 (L)\\
        FrozenBubble Current & 1.0 (L) & 0.93 (L) & 0.88 (L) & 1.0 (L) & 1.0 (L) & 1.0 (L)\\
        \hline
    \end{tabular}
    \label{tab:effect}
\end{table}

%We find that the three algorithms generally perform similarly.
We find that NSGA-II performs similarly to NSGA-III, with the biggest hypervolume in 5 cases for both algorithms.
% We believe that this is due to the fact that the bandwidth of the various patches was the same. 
% NSGA-III was trying to find patches that had better bandwidth fitness due to its reference-point-based selection, but as NSGA-II is only concerned with diversity and not about particular regions of the fitness landscape being occupied. 
% \jp{is this still true given the results in Table 2?}
% It was better able to explore the actual search landscape.
We find that SPEA2 performs best, finding the best fronts in 11 cases.
In general, the different algorithms seem to perform similarly in terms of the best improvements found, as shown in Figure~\ref{fig:exec} and Figure~\ref{fig:mem}.
We find that the caching operators we introduced turned out to be highly effective, appearing in 26\% of improving patches.

We also evaluate the effect size of the improvements found by each of the MO algorithms, as show in Table~\ref{tab:effect}.
We use the Vargha and Delaney A measure~(\cite{vargha}) to calculate the magnitude of the differences between the observations of the NFPs of original applications and the improved versions.
This measure is non-parametric so does assume data is normally distributed.
We find that in all but 8 cases we find large effect sizes, and only find negligible differences in 2 cases.

%\textit{\textbf{RQ3.} \textbf{SPEA2} is the best algorithm for MO-GI for Android.}

\subsection{RQ4: Comparison to SO-GI}

\begin{table}[t]
    \centering
    \caption{Maximum improvements to execution time and memory use found by \tool{}
 using SO-GI (no bandwidth improvements were found).}
    \label{tab:SOGI}
    \begin{tabular}{lrr}
        \toprule
        Application Version & Exec. Time (\%) & Mem. Con. (\%)\\
        \midrule
        PortAuthority 1 & 23.39 & 71.69 \\
        PortAuthority 2 & 21.2 &  53.05 \\
        PortAuthority 3 & 23.13 & 33.76  \\
        PortAuthority 4 & 26.32 & 60.59  \\
        PortAuthority 5 & 28.03 & 59.13 \\
        PortAuthority6 & 24.44 & 24.43 \\
        PortAuthority Current & 29.9 & 9.32 \\ 
        \midrule
        Tower Collector 1 & 16.01 & 30.82 \\
        Tower Collector 2 & 26.92 & 34.61 \\
        Tower Collector Current & 20.9 & 32.43\\
        \midrule
        Gadgetbridge 1 & 29.52 & 31.29   \\
        Gadgetbridge Current  & 26.73 & 5.89 \\
        \midrule
        FosdemComp. 1 & 32.8 & 36.81\\
        FosdemComp. Current & 10.31 & 13.62 \\
        \midrule
        Fdroid 1 & 21.82 & 17.06 \\
        Fdroid 2 & 27.94 & 33.01 \\
        Fdroid Current & 14.14 & 32.18 \\
        \midrule
        LightningBrow. 1 & 28.45 & 8.96 \\
        LightningBro. Current & 23.71 & 32.43 \\
        \midrule
        FrozenBubble 1 & 16.67 & 36.11 \\
        FrozenBubble Current & 19.88 & 4.09\\
        \midrule
    \end{tabular}
\end{table}
Next, we run single-objective genetic improvement on each of our benchmarks.
We measure the effects of the changes found by SO-GI on our other properties.
The results of this evaluation can be found in Table~\ref{tab:SOGI}.
We found improvements to execution time of up to 33\% and memory consumption of up to 72\%.

We find that SO search generally performs better when improving individual properties than multi-objective search.
However, a multi-objective search was capable of finding improvements to both execution time and memory in a similar time as a single-objective search could find improvements to individual properties.
Single-objective search produces results that improve one property in 753 of 1260 cases (21 benchmarks $*$ 20 runs $*$ 3 properties) but in 47\% of these cases, patches are detrimental to another property.

%\textit{\textbf{RQ4.} SO-GI slightly outperforms MO-GI in terms of improving individual properties, but produce patches that are detrimental to other properties \textbf{47\%} of the time.}

\subsection{RQ5: Cost of GI}
\begin{figure}[t]
    \centering
    \includegraphics[scale=0.45]{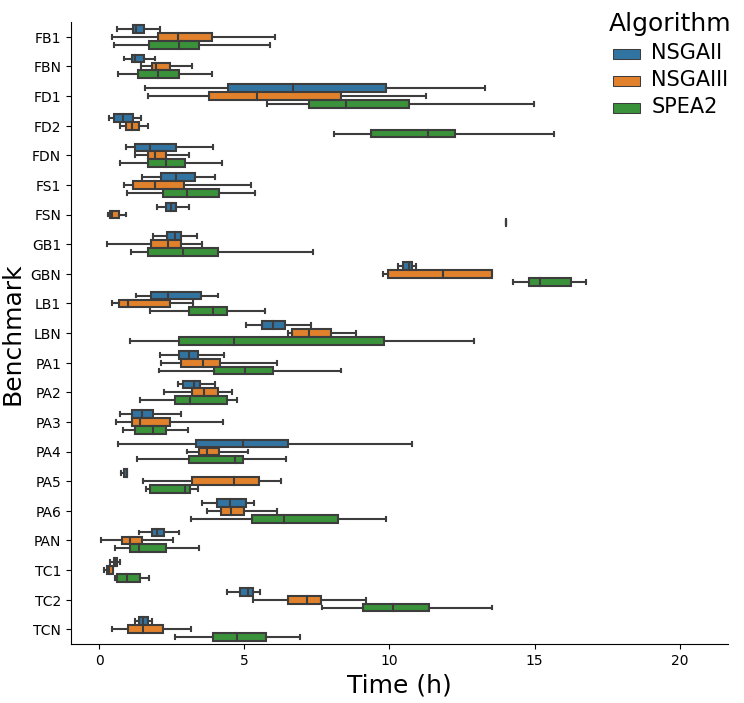}
    \caption{Time taken by \tool{} using different MO algorithms to evolve 10 generations, each with 40 individuals.}
    \label{fig:times}
\end{figure}
In order to evaluate the applicability of our approach, we analyze its cost.
Figure~\ref{fig:times} shows a boxplot of the time taken in hours for our experiments.
We find that the time taken varies a lot between different benchmarks and in some cases even across different runs on the same benchmark.
We find that MO-GI takes between 0.1 hours and 20.6 hours, with a median time across the benchmarks of 2.6 hours.
The main source of variation across the benchmarks is the difference in time taken by the test suites.
In the slowest benchmark, the test suite takes 8 seconds to execute, whereas the quickest one takes 2 seconds.
In the slowest experiments, there were more patches that compiled, rather than instantly failing, further slowing down the experiments.

We find that SO-GI takes longer than MO-GI, with a minimum of 
0.4 hours, a maximum of 19.0 hours, and a median of 3.5 hours.
SO-GI can only find improvements to one property at a time, showing the much-improved efficiency of using MO-GI.
Despite hour-long runtimes, we note that this is a one-off cost.
Given that users consider apps running for 150ms laggy, which might lead to them abandoning an app, we deem the cost of running MO-GI worth it.
%\textit{\textbf{RQ5.} MO-GI takes a median of \textbf{2.6} h, and is \textbf{25\%} faster the SO-GI.}

\subsection{RQ6: Comparison to Linter}
In order to compare our approach to the currently available tooling for improving performance for Android, we run a well-known Linter (PMD) on all of the benchmarks which we improved.
We configured it to provide warnings when any of its performance rules were violated.
We then manually analyzed each of the warnings that it provided, and in the cases where they could be repaired without disrupting the functionality of the application, we repaired them.\looseness=-1

We then measured the performance differences between the repaired and unrepaired versions of the applications.
We found that in our 21 benchmarks, 5 had either no warnings or warnings that could not be repaired without introducing buggy behavior. 
For example, a warning about instantiating an object in a loop could be ``unfixable'' as a reference to each instantiated object is held in an array.
So, moving the instantiation outside of the loop would result in an array with the same reference repeated multiple times.

In all cases where possible, the fixes were easily created and very similar to the examples given in the PMD documentation, and are available in our online repository~(\cite{repo}).

Of the 16 where fixes were possible, only 9 actually offered any improvement.
The maximum improvement to execution time was 4.5\%, while to memory it was 10.42\%, when compared with 35\% and 69\%, respectively, achieved by \tool.
No improvements to bandwidth usage were found.
Only a single one of these patches improved multiple properties, and 6 were detrimental to other properties.
Of those improvements, none had any impact on the bandwidth of the applications.
The linters were, however, significantly quicker than GI, taking a maximum time of 20 minutes to repair the warnings.
However, unlike GI this process is not automatic and requires a developer to be engaged at all times and the improvements found were much smaller than those found by GI.

%\textit{\textbf{RQ6.} \tool{} outperforms linters achieving \textbf{8x} better improvements to execution time and \textbf{7x} better improvements to memory usage.}

%\begin{framed}
%add one-sentence takeaways in boxes, or with different font, space permitting, for each RQ
%\end{framed}
%\jp{the following sentence is too vague, needs rewrite}
%We find that using linters is not an effective way to improve the non-functional properties of Android applications, and MO-GI performs far better.
\begin{table}[t]
    \centering
    \caption{Improvements (\%) from repairing linter warnings, for benchmarks where viable improvements were found.}
    \label{tab:Linter}
    \begin{tabular}{lrrr}
        \toprule
        Application Version & Exec. Time & Mem. Con.  & Time (min.)\\
        \midrule
        PortAuthority 1 & -2.5 & 2.8  & 2\\
        % & e37a1a522a15773710f051d9fff5c0ce68ade5cb  &0 &0 & 0 & 1\\
        % & 3a1329297881aff069cdbc80c92de386ac952d77  &0 &0 & 0 & 1\\
        % & adc73aac9c7dba5c61e1e18a96dfe7dd9712d100  &0 &0 & 0 & 1\\
        PortAuthority 5   &  2.4 & 10.4  & 9\\
        % & a02a0170a38ec257e1f390388e4b5d1414b3cf36  & 0 & 0 & 0 & 4\\
        PortAuthority Current &  0.9 & -2.8 &  1\\
        \midrule
        TowerCollector 
        % 956ea2213c1f7f012d6ab1388536a0c6d5202bd9 & 0& 0& 0 &0\\
        2 & 0 1& 0 &5\\
        TowerCollector Current &  0.0 & 1.9  &7\\
        % \midrule
        % \multirow{2}{*}{Gadgetbridge} & c75362c5ea489247cc00b473a0ef91dbb1cc1569 & 0 & 0 & 0 &4\\
        % & Current & 0 & 0 & 0 & 13\\
        \midrule
        % \multirow{2}{*}{Fosdem Companion} &  b79e29a67c29699b9b8d4ad9c09a3349ce32c59f &0 &0 & 0 &0\\
        % & Current & 0 & 0 & 0 &11\\
        % \midrule
        Fdroid 1 & 4.5 & 0 & 13\\
        % & bf8aa30a576144524e83731a1bad20a1dab3f1bc & 0& 0& 0 &0\\
        Fdroid Current & 2.3 & -0.2 & 9\\
        \midrule
        LightningBrow. 1 & -2.2 & 0.4 &1\\
        LightningBrow. Current &  0.9 & -1.6 &5\\
        \midrule
        FrozenBubble 1 & 3.5 & -0.1 & 20\\
        FrozenBubble Current & -1.6 & 0.4& 15 \\
        \midrule
    \end{tabular}
\end{table}

\section{Threats to Validity}
\label{sec:Threats}
There are a number of threats to the validity of our study.
We discuss these next, including steps we took to mitigate them.\looseness=-1
%We describe here these threats and the steps we took to mitigate these.
%In this section, we will discuss them

%\subsection{Local Testing Only}

The measurements we use for our fitnesses are noisy. 
To mitigate this threat, we repeat each measurement 20 times during search and after the search is complete.
We use the Mann-Whitney U test at the 5\% confidence level to determine whether there is an improvement.
We tested our measurements on known improvements and found that they are consistently detected.\looseness=-1

Furthermore, we use tests to determine whether or not a patch is valid.
This does not guarantee correctness. 
However, the patches produced can undergo the standard code review procedure as any other code being integrated into a project would.
%This code review could also mitigate risks to the code quality that arise from automated patch generation, post-processing tools, such as linters, could be used to assess and improve the code quality of patches
We conducted a manual analysis of all the patches on the Pareto fronts (1753 total), to ensure the improvements reported here do not disturb app functionality.
Through manual analysis, we found that 1352 out of the 1753 best patches found did not disrupt the functionality of the apps, demonstrating the strength of our test suites.
Disruptive patches included the removal of some error handling and the deletion of some components rendered on screen that could not be detected with unit tests.
They would be easily discarded by code review.

Using stochastic search may result in us finding improvements out of sheer luck. %, which are only found on very rare occasions.
In order to avoid this issue, reliably compare different algorithms, and demonstrate generalisability of our approach, we run each of the algorithms tested 20 times on each of our 21 benchmarks.

The search algorithms we use rely on parameters such as mutation and crossover rate. 
The values of these parameters can have an effect on the effectiveness of the algorithms. 
To mitigate this threat, we use the same parameters across all experiments for fair comparison.
We use settings used in previous work that found improvements in software.

We tested our approach on 21 versions of 7 Android apps, which poses a threat to generalisability to other software.
However, these apps are diverse in size and type.
Moreover, we found improvements in current app versions, which were previously undiscovered.
Unfortunately, currently, the big obstacle to wider adoption is test availability. 
For each benchmark, these took us hours to produce.
However, the benefits of testing go beyond the applicability of our approach.
We envision with the development of more fine-grained automated test generation tooling for Android and better testing practices, further benefits of GI can be unlocked.

To mitigate such threats further, we make all our code and results freely available~(\cite{repo}), allowing other researchers and developers to use and extend our tool and validate our work.

\section{Conclusions and Future Work}
\label{sec:Cons}
We propose to use multi-objective genetic improvement (MO-GI) to automatically improve Android apps.
We are the first to apply MO-GI with three objectives to improve software performance and evaluate feasibility of MO-GI for bandwidth and memory use in the Android domain.
To evaluate the effectiveness of the proposed approach we developed \tool{}, which contains 3 MO algorithms and 2 novel cache-based mutation operators.
We have tested \tool{} on 21 benchmarks, targeting runtime, memory, and bandwidth use.
We find improvements to the execution time of up to 35\% and memory consumption of up to 65\%.
However, we find that for the benchmarks we used, our approach cannot find improvements to bandwidth, even though they are within \tool's search space.
Future work could explore the capabilities of large language models for generating non-functional property-improving patches.
Although the techniques currently only perform well on relatively small programs~(\cite{DBLP:journals/corr/abs-2302-07867}), trained on 
source code from programming competitions or puzzles which is short and self-contained~(\cite{DBLP:journals/corr/abs-2105-12655}).
These examples do not contain the complex shared state and interaction with external components that are commonplace in Android apps.
We asked ChatGPT\footnote{\url{https://openai.com/blog/chatgpt/}} to find improvements on our benchmark set with known improvements, but it failed to find any\footnote{All responses are in our repo~(\cite{repo}), in the `Benchmark/ChatGPT' folder.}, while \tool~re-discovered 64\% of those.

%To facilitate uptake and future research, we make our tool and dataset open-source~\cite{repo}.

%We provide the first open-source tool for multi-objective genetic improvement of Android applications, along with a benchmark of 21 versions of Android apps, with tests.

\section{Statements and Declarations}

\textbf{Data availability} All our code and results are available in our repository~\cite{repo}. A permissive open source license will be added upon acceptance.

\noindent\textbf{Funding} This work was supported by EPSRC grant no. EP/P023991/1.

\noindent\textbf{Copyright} For the purpose of open access, the authors have applied a Creative Commons Attribution (CC BY) license to any Accepted Manuscript version arising.

\bibliography{main}

\end{document}